\documentclass[12pt]{article}
\usepackage{bbm}
\def\IZ{\mathbbm Z}

\def\Tr{{\rm Tr}}
\def\and{{\rm and}}
\def\for{{\rm for}}

\def\a{\alpha}
\def\b{\beta}
\def\d{\delta}
\def\g{\gamma}
\def\k{\kappa}
\def\l{\lambda}
\def\s{\sigma}
\def\t{\tau}
\def\om{\omega}

\def\G{\Gamma}
\def\D{\Delta}

\begin{document}
\vspace*{-.6in}
\thispagestyle{empty}
\begin{flushright}
CALT-68-2748\\
\end{flushright}
\baselineskip = 18pt

%\large
%\normalsize

\vspace{1.5in}
{\LARGE
\begin{center}
RECENT PROGRESS IN AdS/CFT\end{center}}
\vspace{.5in}

\begin{center}
John H. Schwarz
\\
\emph{California Institute of Technology\\ Pasadena, CA  91125, USA}
\end{center}
\vspace{1in}

\begin{center}
\textbf{Abstract}
\end{center}
\begin{quotation}
\noindent The study of AdS/CFT (or gauge/gravity) duality has been
one of the most active and illuminating areas of research in string
theory over the past decade. The scope of its relevance and the
insights it is providing seem to be ever expanding. In this talk I
briefly describe some of the attempts to explore how the duality
works for maximally supersymmetric systems.

\end{quotation}

\vfil
\centerline{Contribution to proceedings of the symposium {\it Shifmania}}
%\centerline{\it Crossing the boundaries: Gauge dynamics at strong coupling}
\centerline{celebrating the 60th birthday of M. A. Shifman}

\newpage

\pagenumbering{arabic}

\section{Introduction}

Misha Shifman is one of the most productive (and nicest) theoretical
physicists that I know. It is an honor to have been asked to speak
at this gathering celebrating his 60th birthday. At this stage of my
life, that age seems much younger than it used to.

Recently, in my capacity as an editor of {\it Reviews of Modern
Physics} I persuaded Misha to coauthor a review article entitled
{\em Supersymmetric Solitons and How They Help Us Understand
Non-Abelian Gauge Theories}. I am very pleased to report that this
article has now been published \cite{Shifman:2007ce}. It is already
highly cited.

I have chosen to present an overview of recent progress in AdS/CFT.
An alternative, and more inclusive, name for this subject is {\em
gauge/gravity duality}. There has been progress on various fronts.
In the time available
to me, I will only be able to scratch the surface of this
fascinating subject. There is a great deal of impressive work that I
will not mention at all. Further discussion of some of these topics,
such as the Konishi multiplet,
can be found in Arkady Tseytlin's contribution
\cite{Tseytlin:2009fw}.

\section{Review of Some Basic Facts}

In Maldacena's original paper \cite{Maldacena:1997re}, he proposed
three maximally supersymmetric examples of {AdS/CFT duality}. A basic
indication that the dualities (or equivalences) are plausible
is that the symmetries match. In each case, there is a supergroup,
which describes the isometries of the string theory or M-theory
background geometry. The same supergroup appears as the
superconformal symmetry group of the dual quantum field theory.
Also, the string theory or M-theory solution has $N$ units of flux
threading the sphere factor in the geometry. In fact, the background
configuration corresponds to the near-horizon geometry of $N$
coincident branes, each of which contributes one unit of flux. The
dual conformal field theory, which also depends on the integer $N$,
is the low energy world-volume theory on the branes.

\begin{itemize}
\item {\bf M2-brane Duality:} M-theory on {$AdS_4 \times S^7$}
is dual to a superconformal field theory (SCFT) in three dimensions.
The supergroup is $OSp(8|4)$.

\item {\bf D3-brane Duality:} Type IIB superstring theory on
$AdS_5 \times S^5$ is dual to a SCFT in four dimensions, specifically
${\cal N} =4$ super Yang--Mills (SYM) theory.  The supergroup is
$PSU(2,2|4)$.

\item {\bf M5-brane Duality:} M theory on $AdS_7 \times S^4$
is dual to a SCFT in six dimensions. The supergroup is
$OSp(6,2|4)$.

\end{itemize}

\subsection{The type IIB / ${\cal N} =4$ SYM example}

This by far the most studied, and best understood, example. The $N$
units of flux ($\int_{S^5} F_5 \approx N$) in the superstring
solution correspond to the gauge group $SU(N)$ in the ${\cal N} =4$
super Yang--Mills theory \cite{Brink:1976bc}. The gauge theory has a
well-known large-$N$ topological ('t Hooft) expansion
\cite{'tHooft:1973jz}. The expansion is in powers of $1/N$ for large
$N$ at fixed $\l$, where the 't Hooft parameter is
\begin{equation}
\l = g_{\rm YM}^2 N.
\end{equation}
This expansion corresponds to the loop expansion of the string theory.
One also identifies
\begin{equation}
 R^2/\a' \approx \sqrt{\l} \quad \and \quad g_{\rm s} \approx \l / N ,
\end{equation}
where $R$ is the radius of the $S^5$ and the $AdS_5$. $g_{\rm s}$ is the
string coupling constant determined by the value of the dilaton field.

\subsection{The type IIA / ABJM example}

There has been significant progress in the last couple of years in
understanding the M2-brane duality. The suggestion
\cite{Schwarz:2004yj} that the three-dimensional SCFT should be
Chern--Simons gauge theory was implemented for maximal supersymmetry
(${\cal N} =8$) by Bagger and Lambert \cite{Bagger:2007jr} and by
Gustavsson \cite{Gustavsson:2007vu}. However, their construction
only works for the gauge group $SO(4)$, and it does not provide the
desired dual to M-theory on $AdS_4 \times S^7$.

The correct construction was eventually obtained by Aharony,
Bergman, Jafferis, and Maldacena (ABJM) \cite{Aharony:2008ug}. One
key step in their work was to consider a more general problem:
M-theory on $AdS_4 \times S^7/\IZ_k$, with $N$ units of flux. This
gives 3/4 maximal supersymmetry for $k >2$. Thus, the dual gauge
theory is an ${\cal N} =6$ superconformal Chern--Simons theory in
three dimensions. The appropriate gauge group turns out to be
$U(N)_k \times U(N)_{-k}$, where the subscripts are the levels of
the Chern--Simons terms. The ABJM theory also contains bifundamental
scalar and spinor fields. This theory has a topological expansion,
just like the usual ones in four dimensions, with an 't Hooft
parameter
\begin{equation}
\l = N/k.
\end{equation}
The only unusual feature is that the 't Hooft parameter is rational.
The extension of the supersymmetry from ${\cal N} =6$ to ${\cal N} =8$
for $k=1,2$ is a nontrivial property of the quantum theory.

The orbifold $S^7/\IZ_k$ can be described as a circle bundle over a
$CP^3$ base. The circle has radius $R/k$, where $R$ is the $S^7$
radius. When $k^5 \gg N$, there is a weakly coupled type IIA
superstring interpretation with string coupling constant
\begin{equation}
g_{\rm s} \approx (N/k^5)^{1/4}.
\end{equation}
One then obtains the correspondences
\begin{equation}
R^2/\a' \approx \sqrt{\l} \quad \and \quad g_{\rm s} \approx \l^{5/4}/N,
\end{equation}
which is very similar to the previous duality, but not precisely the
same.

\subsection{AdS energies and conformal dimensions}

The metric of $AdS_{p+2}$ in global coordinates is
\begin{equation}
ds^2[AdS_{p+2}] = d\rho^2 - \cosh^2 \rho \, dt^2
+ \sinh^2 \rho \, ds^2[S^p].
\end{equation}
Here, $ds^2[S^p]$ denote the metric of a unit $p$-dimensional
sphere. Actually, AdS/CFT duality requires taking the covering space
of AdS, which means that the time coordinate $t$ runs from $-\infty$
to $+\infty$.

Witten \cite{Witten:1998qj} and Gubser, Klebanov, Polyakov
\cite{Gubser:1998bc} gave a prescription for relating $n$-point
correlation functions in the gauge theory to corresponding
quantities in the string theory. In the case of two-point functions,
the duality relates the energy $E_A$ of a string state $| A \rangle$
(defined with respect to the global time coordinate $t$)
\begin{equation}
H_{\rm string} | A\rangle = E_A | A \rangle ,
\end{equation}
to the conformal dimensions $\D_A$ of the corresponding
gauge-invariant local operator ${\cal O}_A$ defined by
\begin{equation}
\langle {\cal O}_A (x) {\cal O}_B (y)\rangle \approx
\frac{\d_{AB}}{|x-y|^{2\D_A}}.
\end{equation}

Specifically, the duality requires that
\begin{equation}
\D_A (\l, 1/N) = E_A (R^2/\a', g_{\rm s}).
\end{equation}
The 't Hooft expansion of the dimension of ${\cal O}_A$ is
\begin{equation}
\D_A (\l, 1/N) = \D_A^{(0)} + \sum_{g=0}^\infty \frac{1}{N^{2g}}
\sum_{l=1}^\infty \l^l \D_{l,g} .
\end{equation}
$\D_A^{(0)}$ is the classical (engineering) dimension, and the rest
is called the anomalous dimension.

Almost all studies have focused on the planar approximation, (genus
$g=0$), which is dual to free string theory. This restriction may
make the problem fully tractable, but it is certainly not easy.
After all, it would be an extraordinary achievement to solve an
interacting four-dimensional quantum field theory even in the planar
approximation.

\subsection{Approaches to testing the dualities}

Given that it is not possible to completely solve any of these
theories, the question arises how best to test and explore the
workings of AdS/CFT duality. The most obvious things---matching
symmetries and the dimensions of chiral primary operators---have
been done long ago. One wants to dig deeper. One approach is to
match, as much as possible, energies and dimensions of
fields/operators that are not protected by supersymmetry. It should
be noted, however, that a complete test of the duality would also
require matching three-point correlators, since a conformal field
theory is completely characterized by its two-point and three-point
functions. There has been much less progress on this front.

One approach that has been quite successful is the following. First,
identify tractable examples of classical solutions of the string
world-sheet theory. Next, examine the spectrum of small excitations
about these solutions and compute their energies $E_A$. Finally,
identify the corresponding class of operators in the dual gauge
theory and compute their dimensions $\D_A$ in the planar
approximation. Then compare to $E_A$. One subtlety in this analysis
is that this comparison requires an extrapolation from large $\l$,
where the classical world sheet theory is valid, to small $\l$,
where the gauge theory can be studied perturbatively. Thus, one
needs to identify examples in which this is possible. As we will
see, in practice this has conjectural aspects.

A variant of the preceding procedure is to compare equations that
determine $E_A$ and $\D_A$ rather than the solutions. Approaches
based on integrability and algebraic curves try to obtain equations
of ``Bethe type'' on both sides and to match them. This is a very
active area of research, but I will not be able to review it here.
One important issue is that it is much easier to study the
world-sheet theory when the range of $\s$ is infinite (rather than a
circle). In other words, the string itself is infinite, rather than
a loop. In the gauge theory analysis this corresponds to the
thermodynamic limit of the Bethe equations arising from a spin-chain
analysis. There has been progress recently in extending the
integrability techniques to the compact case \cite{Gromov:2009zb}.
However, the story is
quite technical, and I don't think it is completely settled.

\section{Classical String Solutions}

For the reasons outlined above, we want to identify classical string
solutions in the $AdS_5 \times S^5$ background that can be used to
test the duality. The discussion that follows largely follows an
excellent review article by Plefka \cite{Plefka:2005bk}. Other
useful reviews include \cite{Tseytlin:2003ii,Kristjansen:2009}.

The bosonic part of the string world-sheet
action has six cyclic coordinates:
\begin{equation}
(t,\varphi_1 , \varphi_2; \phi_1,\phi_2,\phi_3),
\end{equation}
where the first three coordinates pertain to $AdS_5$ and the second
three to $S^5$. Specifically, we parametrize $S^5$ as follows:
\begin{equation}
ds^2 (S^5) = d\g^2 + \cos^2\g \, d\phi_3^2 + \sin^2\g \, ds^2(S^3),
\end{equation}
where
\begin{equation}
ds^2(S^3) = d\psi^2 +\cos^2\psi \, d \phi_1^2 + \sin^2\psi \, d\phi_2^2.
\end{equation}
Associated to these cyclic coordinates one has conserved charges
\begin{equation}
(E, S_1, S_2; J_1, J_2, J_3).
\end{equation}
$E$ is the energy and the other five charges are angular momenta.

One much-studied class of string solutions involves a line up the
center of $AdS_5$, described by $\rho = 0$ and $t =\k \t$, where
$\k$ is a constant and $\t$ is the world-sheet time coordinate.
These configurations have $S_1 =S_2 =0$.

\subsection{Point-particle solutions}

The simplest solution is a point particle (collapsed string)
encircling the sphere. In addition to $\rho = 0$ and $t =\k \t$,
this is described by
\begin{equation}
\g =\pi/2, \quad  \phi_1 =\k\t,  \quad \psi=0.
\end{equation}
This has $J_2 =J_3 =0$.

The quantum excitations of this solution have energies that can be
expanded in powers of $1/J$ for large $J=J_1$, where
\begin{equation}
\k = J/\sqrt{\l}
\end{equation}
is held fixed. This is equivalent to the Berenstein, Maldacena
Nastase (BMN) analysis of strings in a plane-wave background
\cite{Berenstein:2002jq}. One obtains
\begin{equation}
E-J \approx E_2(\k) + \frac{1}{J} E_4(\k) + \ldots
\end{equation}

The exact BMN result is
\begin{equation}
E_2 = \sum_{n= -\infty}^{\infty} \sqrt{n^2 + \k^2} N_n,
\end{equation}
where $N_n = \sum_{i=1}^8 \a_n^{i\dagger} \a_n^i + {\rm fermions}$
is expressed in terms of ordinary oscillators
\begin{equation}
[\a_m^i,\a_n^{j\dagger}] = \d^{ij} \d_{mn}.
\end{equation}
The level matching condition is $\sum n N_n =0$.

The BMN paper proposed a scaling rule, known as BMN scaling, which
predicts agreement with the anomalous dimensions of operators in the
dual gauge theory, even though one calculation is valid for large
$\l$ and the other for small $\l$. In other words, their scaling
hypothesis, if valid, would justify the extrapolation from small
$\l$ to large $\l$. In fact, it turns out that $E_2$ agrees
perfectly, but agreement for $E_4$ breaks down at three loops
\cite{Callan:2003xr}. This is not a problem for AdS/CFT duality,
only for the BMN scaling conjecture.\footnote{Perhaps it would be
more fair to say that the BMN scaling conjecture was made for the
plane-wave limit only, which corresponds to $E_2$; what fails is an
attempt to generalize the scaling conjecture beyond that.}

\subsection{Spinning string solutions}

A class of interesting generalizations of the preceding solution
describes circular or folded strings that are extended on the $S^3
\subset S^5$. These have $t = \k\t$, $\rho = 0$, and $\g = \pi/2$,
as before. But now one takes
\begin{equation}
\phi_1 = \om_1 \t, \quad \phi_2 =
\om_2 \t, \quad \psi = \psi(\s).
\end{equation}
For these choices, the string equation of motion gives
\begin{equation}
\psi'' + \om_{21}^2  \sin\psi \cos\psi =0,
\end{equation}
where $\om_{21}^2 = \om_2^2 - \om_1^2$. This is the well-known
pendulum equation.

This equation has a first integral
\begin{equation}
\psi' = \om_{21} \sqrt{q-\sin^2\psi}, \quad q= (\k^2
-\om_1^2)/\om_{21}^2.
\end{equation}
The solution for  $q<1$, which involves the elliptic integrals
$E(q)$ and $K(q)$, describes a folded string. It corresponds to a
pendulum that oscillates back and forth. The solution for $q>1$,
which involves the elliptic integrals $E(q^{-1})$ and $K(q^{-1})$,
describes a circular string. It corresponds to a pendulum that goes
round and round. In the classical limit, the energy has the form
\begin{equation}
E = \sqrt{\l} F( J_1/\sqrt{\l}, J_2/\sqrt{\l}).
\end{equation}

\subsection{Dual gauge theory analysis}

This string theory result can be extrapolated to small $\l$ and
compared to the dual gauge theory.  The operators that carry $J_1,
J_2$ charges have the form
\begin{equation}
{\cal O}_\a^{J_1, J_2} = \Tr\left( Z^{J_1} W^{J_2}\right) + \ldots
\end{equation}
where $Z$ and $W$ are complex scalar fields in the adjoint of $SU(N)$.
The additional terms denoted by dots involve different orderings of the
$Z$s and $W$s. Such a trace can be viewed as a ring configuration of an
$S=1/2$ quantum spin chain, where $W$ corresponds to spin up and $Z$
corresponds to spin down.

The conformal dimensions of operators ${\cal O}_\a^{J_1, J_2}(x)$
with these charges are eigenvalues of the dilatation operator
\begin{equation}
{\cal D} {\cal O}_\a^{J_1, J_2}(x) = \sum _\b D_{\a\b} {\cal
O}_\b^{J_1, J_2}(x).
\end{equation}
In the planar one-loop approximation the equations are precisely
those of a ferromagnetic Heisenberg spin chain, which is a
well-known integrable system, whose Hamiltonian is proportional to
\begin{equation}
{\cal H} = \sum _{i=1}^J \left(\frac{1}{4} -
\vec\s_i\cdot\vec\s_{i+1}\right).
\end{equation}
This can be solved using Bethe ansatz techniques, thereby obtaining
conformal dimensions that can be compared (successfully) with
energies of the corresponding string solutions. Higher-order terms,
which correspond to more complicated spin-chain Hamiltonians, have
also been studied.

\subsection{Strings spinning in AdS}

Another interesting class of classical string solutions are ones in
which the string position is extended in the AdS space and a
point moving on the sphere (rather than the other way around).
The first example of this type is the
straight folded string rotating in $AdS_3 \subset AdS_5$
\cite{Gubser:2002tv}. One finds that for large $S$
\begin{equation}
E= 2\G(\l) \log S + O(S^0),
\end{equation}
where
\begin{equation}
\G(\l) = \frac{\sqrt{\l}}{2\pi} + O(\l^0) \quad \for \quad \l \gg 1.
\end{equation}

The dual gauge theory operators are
\begin{equation}
\Tr (D_+^{s_1} Z \, D_+^{s_2} Z) \quad s_1 + s_2 =S.
\end{equation}
Their anomalous dimensions take the same form as the energy with
\begin{equation}
\G(\l) = \frac{\l}{4\pi^2} + O(\l^2)  \quad \for \quad \l \ll 1.
\end{equation}
In order to compare these, one needs a procedure to extrapolate
between small and large $\l$. In fact, an exact formula for the {\em
cusp anomalous dimension} $\G(\l)$ has been deduced using the
assumption of exact integrability \cite{Beisert:2006ez}. It passes
all tests and is likely to be correct.

The generalization of this duality to twist $J$ operators, which
have the form
\begin{equation}
\Tr (D_+^{s_1} Z \, D_+^{s_2} Z \ldots D_+^{s_J} Z) \quad {\rm
where} \quad\sum s_l=S,
\end{equation}
has been explored by Dorey and Losi \cite{Dorey:2008vp}. They
computed the corresponding conformal dimensions using an $SL(2)$
spin chain model. For large $S$ the correspond classical string
solutions are {\em spiky strings} with $J$ cusps. The duality
predictions are verified to the extent that they have been explored.

\section{Conclusion}
There has been a lot of progress in testing AdS/CFT in various
special cases for maximally supersymmetric theories. Much of this
progress has exploited the integrability of the string world-sheet
theory on the one hand and the integrability of various spin-chain
models that arise in studies of the dual gauge theory in the planar
approximation on the other hand.

I have described classical string solutions and the dual gauge
theory operators that are relevant to the D3-brane duality. There
has also been very interesting work exploring analogous
constructions for the M2-brane duality following the discovery of
the ABJM theory. Much less is known about the M5-brane theory,
though there has been significant progress when two of the
dimensions wrap a Riemann surface \cite{Gaiotto:2009we,Gaiotto:2009gz}.

\section*{Acknowledgments}

This work was supported in part by the U.S. Dept. of
Energy under Grant No. DE-FG03-92-ER40701.


\newpage

\begin{thebibliography}{99}

%\cite{Shifman:2007ce}
\bibitem{Shifman:2007ce}
  M.~Shifman and A.~Yung,
  ``Supersymmetric Solitons and How They Help Us Understand Non-Abelian Gauge
  Theories,''
  Rev.\ Mod.\ Phys.\  {\bf 79}, 1139 (2007)
  [arXiv:hep-th/0703267].
  %%CITATION = RMPHA,79,1139;%%

%\cite{Tseytlin:2009fw}
\bibitem{Tseytlin:2009fw}
  A.~A.~Tseytlin,
  ``Quantum Strings in AdS5 x S5 and AdS/CFT Duality,''
  arXiv:0907.3238 [hep-th].
  %%CITATION = ARXIV:0907.3238;%%

%\cite{Maldacena:1997re}
\bibitem{Maldacena:1997re}
  J.~M.~Maldacena,
  ``The Large N Limit of Superconformal Field Theories and Ssupergravity,''
  Adv.\ Theor.\ Math.\ Phys.\  {\bf 2}, 231 (1998)
  [Int.\ J.\ Theor.\ Phys.\  {\bf 38}, 1113 (1999)]
  [arXiv:hep-th/9711200].
  %%CITATION = IJTPB,38,1113;%%

%\cite{Brink:1976bc}
\bibitem{Brink:1976bc}
  L.~Brink, J.~H.~Schwarz and J.~Scherk,
  ``Supersymmetric Yang--Mills Theories,''
  Nucl.\ Phys.\  B {\bf 121}, 77 (1977).
  %%CITATION = NUPHA,B121,77;%%

%\cite{'tHooft:1973jz}
\bibitem{'tHooft:1973jz}
  G.~'t Hooft,
  ``A Planar Diagram Theory for Strong Interactions,''
  Nucl.\ Phys.\  B {\bf 72}, 461 (1974).
  %%CITATION = NUPHA,B72,461;%%

%\cite{Schwarz:2004yj}
\bibitem{Schwarz:2004yj}
  J.~H.~Schwarz,
  ``Superconformal Chern--Simons Theories,''
  JHEP {\bf 0411}, 078 (2004)
  [arXiv:hep-th/0411077].
  %%CITATION = JHEPA,0411,078;%%

%\cite{Bagger:2007jr}
\bibitem{Bagger:2007jr}
  J.~Bagger and N.~Lambert,
  ``Gauge Symmetry and Supersymmetry of Multiple M2-Branes,''
  Phys.\ Rev.\  D {\bf 77}, 065008 (2008)
  [arXiv:0711.0955 [hep-th]].
  %%CITATION = PHRVA,D77,065008;%%

%\cite{Gustavsson:2007vu}
\bibitem{Gustavsson:2007vu}
  A.~Gustavsson,
  ``Algebraic Structures on Parallel M2-Branes,''
  Nucl.\ Phys.\  B {\bf 811}, 66 (2009)
  [arXiv:0709.1260 [hep-th]].
  %%CITATION = NUPHA,B811,66;%%

%\cite{Aharony:2008ug}
\bibitem{Aharony:2008ug}
  O.~Aharony, O.~Bergman, D.~L.~Jafferis and J.~M.~Maldacena,
  ``$N=6$ Superconformal Chern--Simons-Matter Theories, M2-Branes and Their
  Gravity Duals,''
  JHEP {\bf 0810}, 091 (2008)
  [arXiv:0806.1218 [hep-th]].
  %%CITATION = JHEPA,0810,091;%%

%\cite{Witten:1998qj}
\bibitem{Witten:1998qj}
  E.~Witten,
  ``Anti de Sitter Space and Holography,''
  Adv.\ Theor.\ Math.\ Phys.\  {\bf 2}, 253 (1998)
  [arXiv:hep-th/9802150].
  %%CITATION = 00203,2,253;%%

%\cite{Gubser:1998bc}
\bibitem{Gubser:1998bc}
  S.~S.~Gubser, I.~R.~Klebanov and A.~M.~Polyakov,
  ``Gauge Theory Correlators from Non-critical String Theory,''
  Phys.\ Lett.\  B {\bf 428}, 105 (1998)
  [arXiv:hep-th/9802109].
  %%CITATION = PHLTA,B428,105;%%

%\cite{Gromov:2009zb}
\bibitem{Gromov:2009zb}
  N.~Gromov, V.~Kazakov and P.~Vieira,
  ``Exact AdS/CFT Spectrum: Konishi Dimension at any Coupling,''
  arXiv:0906.4240 [hep-th].
  %%CITATION = ARXIV:0906.4240;%%

%\cite{Plefka:2005bk}
\bibitem{Plefka:2005bk}
  J.~Plefka,
  ``Spinning Strings and Integrable Spin Chains in the AdS/CFT
  Correspondence,''
  Living Rev.\ Rel.\  {\bf 8}, 9 (2005)
  [arXiv:hep-th/0507136].

%\cite{Tseytlin:2003ii}
\bibitem{Tseytlin:2003ii}
  A.~A.~Tseytlin,
  ``Spinning Strings and AdS/CFT Duality,''
  arXiv:hep-th/0311139.
  %%CITATION = HEP-TH/0311139;%%

%\cite{Kristjansen:2009}
\bibitem{Kristjansen:2009}
C. Kristjansen, M. Staudacher, and A. Tseytlin, eds., {\it Special
Issue on Integrability and the AdS/CFT Correspondence}, J. Phys. A
{\bf 42}, 250301 (2009).

%\cite{Berenstein:2002jq}
\bibitem{Berenstein:2002jq}
  D.~E.~Berenstein, J.~M.~Maldacena and H.~S.~Nastase,
  ``Strings in Flat Space and pp Waves from $N = 4$ Super Yang Mills,''
  JHEP {\bf 0204}, 013 (2002)
  [arXiv:hep-th/0202021].
  %%CITATION = JHEPA,0204,013;%%

%\cite{Callan:2003xr}
\bibitem{Callan:2003xr}
  C.~G.~Callan, H.~K.~Lee, T.~McLoughlin, J.~H.~Schwarz, I.~Swanson and X.~Wu,
  ``Quantizing String Theory in AdS(5) x S**5: Beyond the pp-Wave,''
  Nucl.\ Phys.\  B {\bf 673}, 3 (2003)
  [arXiv:hep-th/0307032].
  %%CITATION = NUPHA,B673,3;%%

%\cite{Gubser:2002tv}
\bibitem{Gubser:2002tv}
  S.~S.~Gubser, I.~R.~Klebanov and A.~M.~Polyakov,
  ``A Semi-Classical Limit of the Gauge/String Correspondence,''
  Nucl.\ Phys.\  B {\bf 636}, 99 (2002)
  [arXiv:hep-th/0204051].
  %%CITATION = NUPHA,B636,99;%%

%\cite{Beisert:2006ez}
\bibitem{Beisert:2006ez}
  N.~Beisert, B.~Eden and M.~Staudacher,
  ``Transcendentality and Crossing,''
  J.\ Stat.\ Mech.\  {\bf 0701}, P021 (2007)
  [arXiv:hep-th/0610251].
  %%CITATION = JSTAT,0701,P021;%%

%\cite{Dorey:2008vp}
\bibitem{Dorey:2008vp}
  N.~Dorey and M.~Losi,
  ``Spiky Strings and Spin Chains,''
  arXiv:0812.1704 [hep-th].
  %%CITATION = ARXIV:0812.1704;%%

%\cite{Gaiotto:2009we}
\bibitem{Gaiotto:2009we}
  D.~Gaiotto,
  ``$N=2$ Dualities,''
  arXiv:0904.2715 [hep-th].
  %%CITATION = ARXIV:0904.2715;%%

%\cite{Gaiotto:2009gz}
\bibitem{Gaiotto:2009gz}
  D.~Gaiotto and J.~M.~Maldacena,
  ``The Gravity Duals of $N=2$ Superconformal Field Theories,''
  arXiv:0904.4466 [hep-th].
  %%CITATION = ARXIV:0904.4466;%%


\end{thebibliography}
\end{document}